\documentclass[final,square,comma,sort&compress,5p,times,nofootinbib,twocolumn]{elsarticle}
\usepackage{color}
\usepackage{amssymb}
\usepackage[skip=1.0\baselineskip]{caption}
\usepackage{lineno,hyperref}
\modulolinenumbers[2]
\journal{Physics Letters B}

\usepackage{graphicx}
\usepackage{adjustbox}
\usepackage{dcolumn}
\usepackage{bm}
\usepackage{multirow}
\usepackage{amsmath}

\usepackage{amsmath}
\usepackage{mathrsfs}

\usepackage[final]{changes}

\begin{document}

\begin{frontmatter}

\title{ Structure of $^{29}$F in the Rotation-aligned Coupling Scheme \\ of the Particle-Rotor Model }

\author{A.~O.~Macchiavelli, P.~Fallon, H.~L.~Crawford, C.~M.~Campbell, R.~M.~Clark, M.~Cromaz,  M.~D.~Jones, I.~Y.~Lee, and M.~Salathe}
\address{Nuclear Science Division, Lawrence Berkeley National Laboratory, Berkeley, CA 94720, USA}
%\affiliation{$^{2}$Department of Physics and Astronomy, Ohio University, Athens, OH 45701, USA}
\date{\today}
 
\begin{abstract}
Recent results from RIKEN/RIBF on the low-lying level structure of $^{29}$F\deleted{~\cite{Pieter}} are interpreted \replaced{within}{with} the Particle-Rotor Model.  We show that the experimental 
data can be understood in the Rotation-aligned Coupling Scheme, with the $5/2^+$ ground state as the bandhead of a decoupled band.  In this picture,  the energy of
the observed $1/2_1^+$ state correlates strongly with the rotational energy of the core and provides an estimate of the $2^+$ energy in $^{28}$O.  Our analysis suggest 
a moderate deformation,  $\epsilon_2 \sim 0.17$,  and places the $2^+$ in $^{28}$O at $\sim$ 2.4 MeV.   
\end{abstract}

\end{frontmatter}
%\pacs{21.10.Hw, 27.30.+t, 21.10.Jx, 25.55.Hp}

%\maketitle

%\section{Introduction}
%\twocolumn
%\linenumbers

The structure of exotic neutron-rich nuclei is one of the main science drivers  in contemporary  \replaced{nuclear physics}{Nuclear Physics} research.
Our current knowledge of nuclear structure, towards the driplines, has clearly established that the paradigm of 
magic numbers and doubly magic nuclei as we know \added{it} near stability changes across the nuclear landscape~\cite{Sor08}.
It is expected that these changes in the underlying single-particle structure are intimately related to specific aspects of the 
effective nuclear force, specifically to its tensor and three-body \added{(or higher)} components.

Thus, a detailed mapping of shell evolution and collectivity at the limits of isospin becomes a key element 
to understand the atomic nucleus and all of its many-body intricacies.  The Islands of Inversion at  $N$=8, 20, and 40 provide 
dramatic examples\deleted{of these} with the underlying physics mechanism driven by the important role of the neutron-proton force~\cite{Sor08, Poves, Ots01, Heyde1}.
The effect of isospin on the monopole average of the central and tensor components of the force affects the neutron effective single-particle energies (ESPEs) 
in such a way that expected shell closures are quenched, opening the door for the collective degrees of freedom to become relevant in the low-lying excitation 
spectra of these systems, where single-particle excitations were anticipated.

The study of odd-$A$ and odd-odd nuclei has \deleted{been}traditionally \added{been} a tool of choice to disentangle the competition of  single-particle and
collective motion insofar as they can be \added{regarded}, at least a priori, \deleted{regarded} as one or two valence nucleon\added{(s)} coupled to a core.

As nicely discussed in Ref.~\cite{Pieter}, a very appealing region to study is that near $^{28}$O, which is today accessible \deleted{both} experimentally and \added{also}
theoretically with state-of-the-art large scale Shell Model calculations.  The well known magic numbers, $Z$=8 and $N$=20, 
would anticipate  $^{28}$O as \deleted{of}a doubly magic nucleus, however, \replaced{the recent}{their} study of $^{29}$F carried out at RIKEN/RIBF suggests otherwise~\cite{Pieter}.
As noted,  the relatively low transition energy of the $1/2_1^+$ state in $^{29}$F (1.08~MeV) largely disagrees with Shell Model predictions restricted to the $sd$ model
space, $\approx$ 3.5 MeV.  Based on their calculations using the SDPF-M effective interaction, the authors determined that the $N$= 20 shell gap is quenched for $^{29}$F and
concluded:  {\sl "... \deleted{thus} extending the Island of Inversion to isotopes with proton number $Z$= 9."}

In this work we follow-up on \replaced{this}{their} conclusion and ask \deleted{ourselves} whether the observed structure in $^{29}$F is amenable \replaced{to}{of} a description in terms of
a collective picture~\cite{BM}, within the framework of the Particle-Rotor Model (PRM)~\cite{Larsson, Rag}.

%\section{The Decoupled Limit}
Considering $^{28}$O as our core, an inspection of a Nilsson diagram~\cite{Sven} suggests that the well bound odd proton will occupy the single-$j$ multiplet 
originating from the $d_{5/2}$ orbit, namely the levels  $\frac{1}{2}[220]$, $\frac{3}{2}[211]$, and $\frac{5}{2}[202]$, with its Fermi energy at the $\Omega=1/2$.\\

\noindent
The Hamiltonian of the system can be written as~\cite{Larsson,Rag}:
\begin{equation}
H=      E_\Omega +   \frac{\hbar^{2}}{2\mathscr{I}}I(I+1) + H_C
\label{eq:eq1}
\end{equation}

\noindent
Here $E_\Omega$ relates to the intrinsic level energies,  $\mathscr{I}$ is the core moment-of-inertia, and $H_C$ the Coriolis coupling term given by

\begin{equation}
H_C=   - \frac{\hbar^{2}}{2\mathscr{I} }(I_+j_-+I_-j_+)
\label{eq:eq2}
\end{equation}
where $I_\pm$ and  $j_\pm$ the ladder operators for the total and single particle angular momenta respectively. \\

Given the conditions above and for small to moderate deformations ($\epsilon_2 \sim 0.15$) the Coriolis matrix elements\footnote[1]{We have not used any explicit attenuation of these matrix elements in the PRM calculations.} ($\sim \hbar^2Ij/\mathscr{I}$) dominate over the intrinsic level spacings,
($\Delta E_{\Omega,\Omega\pm1} \sim \hbar\omega_0\epsilon_2$), and
a rotation-aligned coupling limit is anticipated~\cite{frank1,frank}. In this case, the lowest-lying \added{({\sl yrast})} band \deleted{({\sl yrast})} has spins $I=j,~j+2,~j+4,~...$,  and energy spacings equal to that of the core; this type of 
band is referred to as a decoupled band. Specific to our case, the ground state is then predicted to be $5/2^+$, for which a geometrical picture is shown on the left panel of Fig. 1.

\begin{figure}[htbp]
\centering
\includegraphics[width=8.0cm]{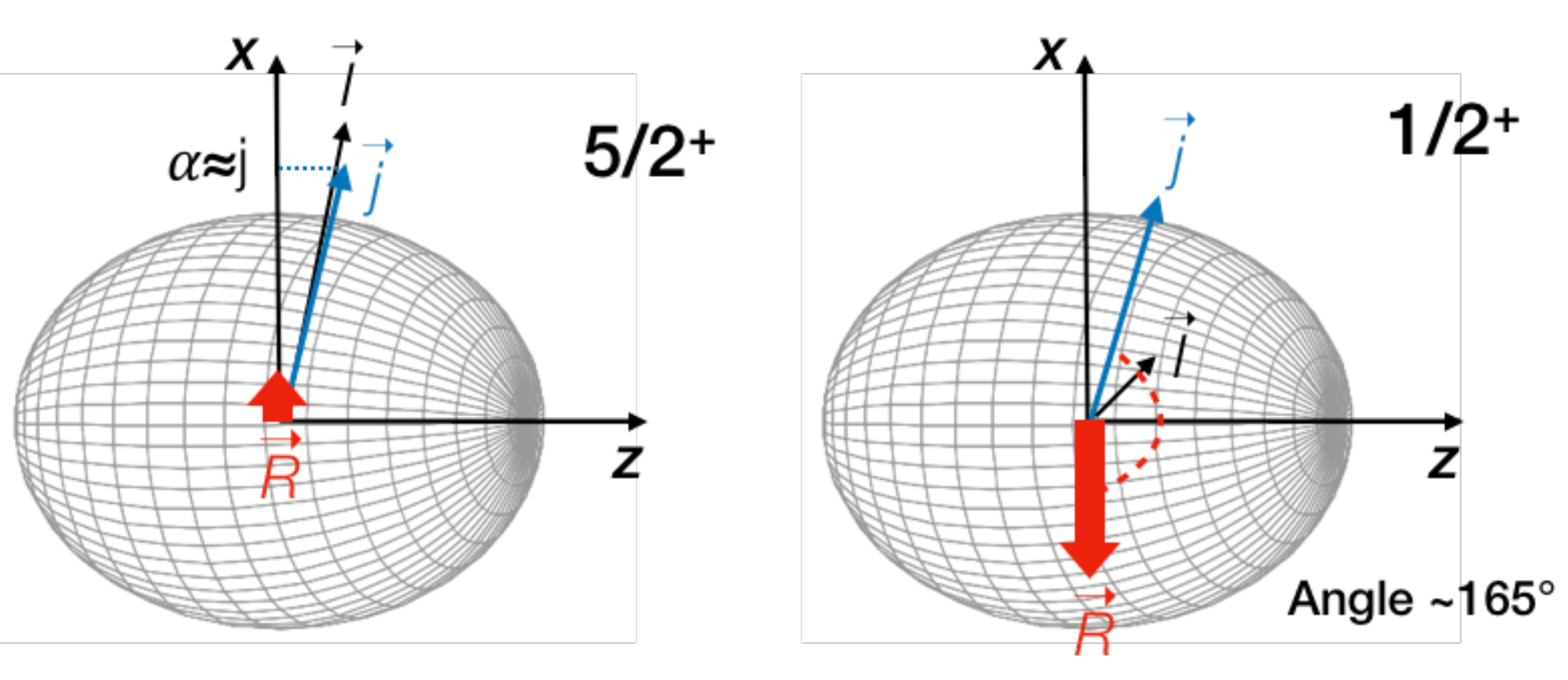}
\caption{ Schematic diagram to illustrate the vector coupling for the lowest states in $^{29}$F. The black arrows represent the total momentum, $\vec{I}$, while the blue arrows represent $\vec{j}$ and the red arrows the core rotation $\vec{R}.$}
\label{fig:schematic}
\end{figure}

As illustrated on the right panel,  the excited $1/2^+_1$ state must have,  by necessity,  anti-parallel coupling of $\vec{j}$ with the core rotation, $\vec{R}$.\footnote[2]{The angle can be estimated from the semi-classical expression:  $ cos~\theta = \frac{1}{2}(I(I+1) - R(R+1) -j(j+1))/\sqrt{R(R+1)j(j+1))}$,  giving $\theta \approx 165^\circ$.}
Therefore it follows that in the decoupled limit ($\epsilon_2 \rightarrow 0$) the energy of the $1/2^+_1$ state\deleted{s} with respect to the
ground state is proportional to the rotational energy of the core, $E_{2^+}(core)$,
%\begin{equation}
%%E_{1/2^+}-E_{5/2^+} \propto E_{2^+}(core)/3
%\end{equation}
and provides\added{, in the case of $^{29}$F,} a proxy for the $2^+$ energy in $^{28}$O.  Since 
the splitting of the Nilsson multiplet is proportional to the quadrupole deformation, $\sim\hbar\omega_0\epsilon_2$, we  expect a trade-off between $E_{2^+}(core)$ vs. $\epsilon_2$, and thus a range of possible solutions matching the experimental results. \\
%\begin{figure}[htbp]
%\centering
%\includegraphics[width=8.0cm]{Fig2b.pdf}
%\caption{  Energies of $1/2^+_1$ state in $^{29}$F as a function of the $2^+$ energy of the core for some values of the quadrupole deformation $\epsilon_2$, compared to the experimental result. The arrow indicates our adopted solution (see text).}
%\label{fig:schematic}
%\end{figure}the level mixing induced by $H_C$ depends on the ratio $E_{2^+}(core)/(E_{\Omega_1} - E_{\Omega_2})$, and

\noindent
In Fig.~2 we  compare \deleted{the} PRM solutions as a function of the deformation, to a calculation of the core energy given by  $E_{2^+}= 3\hbar^{2}/\mathscr{I}$ with  the moment of inertia calculated using the Migdal formula~\cite{Migdal,BM2,Bengtsson},
\begin{equation}
\mathscr{I}=\frac{\mathscr{I}_{rigid}}{(1+(\frac{2\Delta}{\hbar\omega_0\epsilon_2})^2)^{3/2}}
\end{equation}
\noindent
using an isospin dependent expression of the pairing gap, $\Delta$, from Ref.~\cite{gregers}, adjusted to this region of the nuclear chart.

The values $E_{2^+}$ and $\epsilon_2$ where the two curves intersect defines a consistent solution to the problem.  To estimate an uncertainty in the adopted solution we take into account  an uncertainty of $\sim \pm$ 13\% in $\Delta$ entering the calculation of $\mathscr{I}$~\cite{aom}, and obtain the blue shaded band, from which $E_{2^+} \approx 2.\replaced{4}{3}\replaced{\pm0.2}{^{+0.2}_{-0.1}}$~MeV and  $\epsilon_2 \approx 0.17^{+0.15}_{-0.2}$. \\
\begin{figure}[htbp]
\centering
\includegraphics[width=9.0cm]{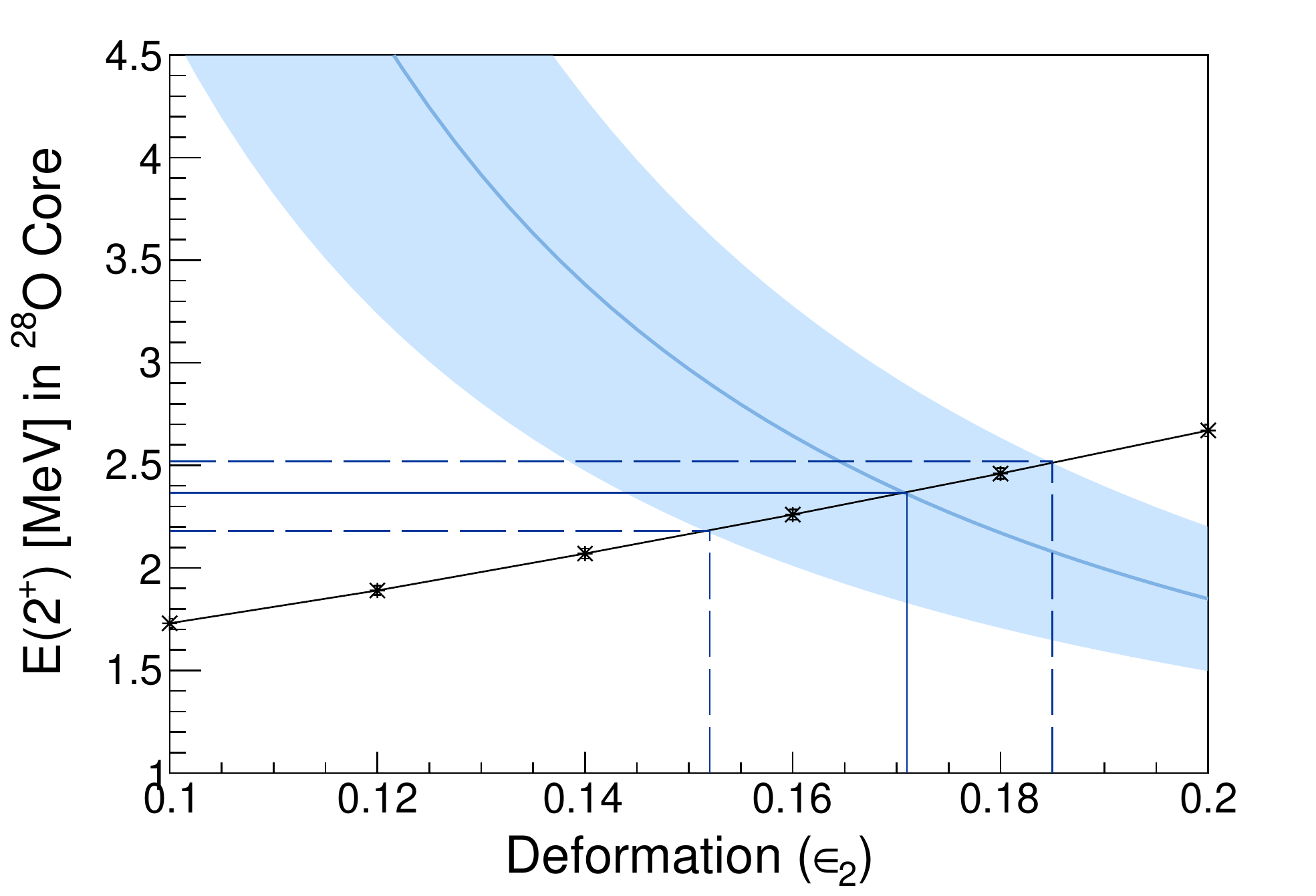}
\caption{ The $E_{2^+}(core)$  required to: 1) reproduce the energy of the $1/2^+_1$ state, 1.08~MeV,  as a function of deformation (black line) and 2) calculated with the Migdal formula (blue line) using a  pairing gap, $\Delta$=1.5 MeV. The shaded band is an estimate of the 
theoretical error in the calculation of $\mathscr{I}$.}
\end{figure}

\noindent
A given state, $ | I,\alpha\rangle$,  of angular momentum $I$ and projection $\alpha$ onto the rotation axis (x-axis) has a wavefunction of the form:
\begin{equation}
| I,\alpha\rangle =  \sum_{\Omega=1/2}^{5/2} C^\alpha_{I\Omega}|I,\Omega\rangle
\label{eq:eq1}
\end{equation}
It can be shown~\cite{frank1,frank}, that in the rotation-aligned coupling limit,  the amplitudes are given by the  
Wigner $d$-function evaluated at $\pi/2$, the angle between the symmetry and rotation axes:
\begin{equation}
C^\alpha_{I\Omega}=d^j_{\alpha,\Omega}(\pi/2)
\label{eq:eq1}
\end{equation}

\begin{table*}[ht]
\centering
%%\begin{adjustbox}{max width=\textwidth}

\caption{ PRM results for the low-lying levels of $^{29}$F.  The lowest two states have been observed experimentally. Magnetic moments have been calculated with no-quenching of $g_s$, and $g_R=Z/A$. (In  boldface we indicate the {\sl yrast} band members).}

{\renewcommand{\arraystretch}{1.5}
%\scalebox{0.9}{
\begin{tabular}{c|c|c|c|c|c|c|c|c}
\hline\hline
State &  Energy  & ~$\langle R \rangle$ ~&$E_{rot}$ &  $\langle I_z \rangle$ & $\langle \vec{I} \cdot \vec{j} \rangle/ | I |$ &  $\langle \vec{R} \cdot \vec{j} \rangle/ | R |$ & Magnetic Moment  & Quadrupole Moment \\
         & [MeV]     &                                     &   [MeV]   &                                    &                                                                  &                                                                      &   [$\mu_N$]             &    [eb]\\
\hline 
{\bf 5/2}$^+$& 0 & 0.67 &0.43&0.08&2.65 & -0.21&4.6& -0.06\\
$1/2^+$& 1.08 & 1.84&2.00&0.5 &1.83&-2.65 &  2.4& 0\\
\hline          
$3/2^+$& 2.2 &  2.01& 2.32&-1.04&1.58&-2.19&2.5& 0.026\\
{\bf 9/2}$^+$&2.6 &2.28 &2.9&0.05& 2.55& 1.67&  5.3& -0.1\\
$7/2^+$& 3.2&  2.1& 2.50& 0.48&2.25& 0.12 &4.1& -0.024\\
                  \hline\hline
\end{tabular}
}
\label{Table 2}
\end{table*}

\noindent
For our adopted PRM solution, the square amplitudes for the lowest two states, $5/2^+ $ and $9/2^+$  of the {\sl yrast} band ($\alpha=j=5/2$) are given  in Fig.~3 where they are compared to the limit above
showing that the structure can be interpreted, indeed, as a decoupled band. Some geometrical and spectroscopic properties of the calculated low-lying levels are summarized in Table~1.\\
\begin{figure}[htbp]
\centering
\includegraphics[width=9.0cm]{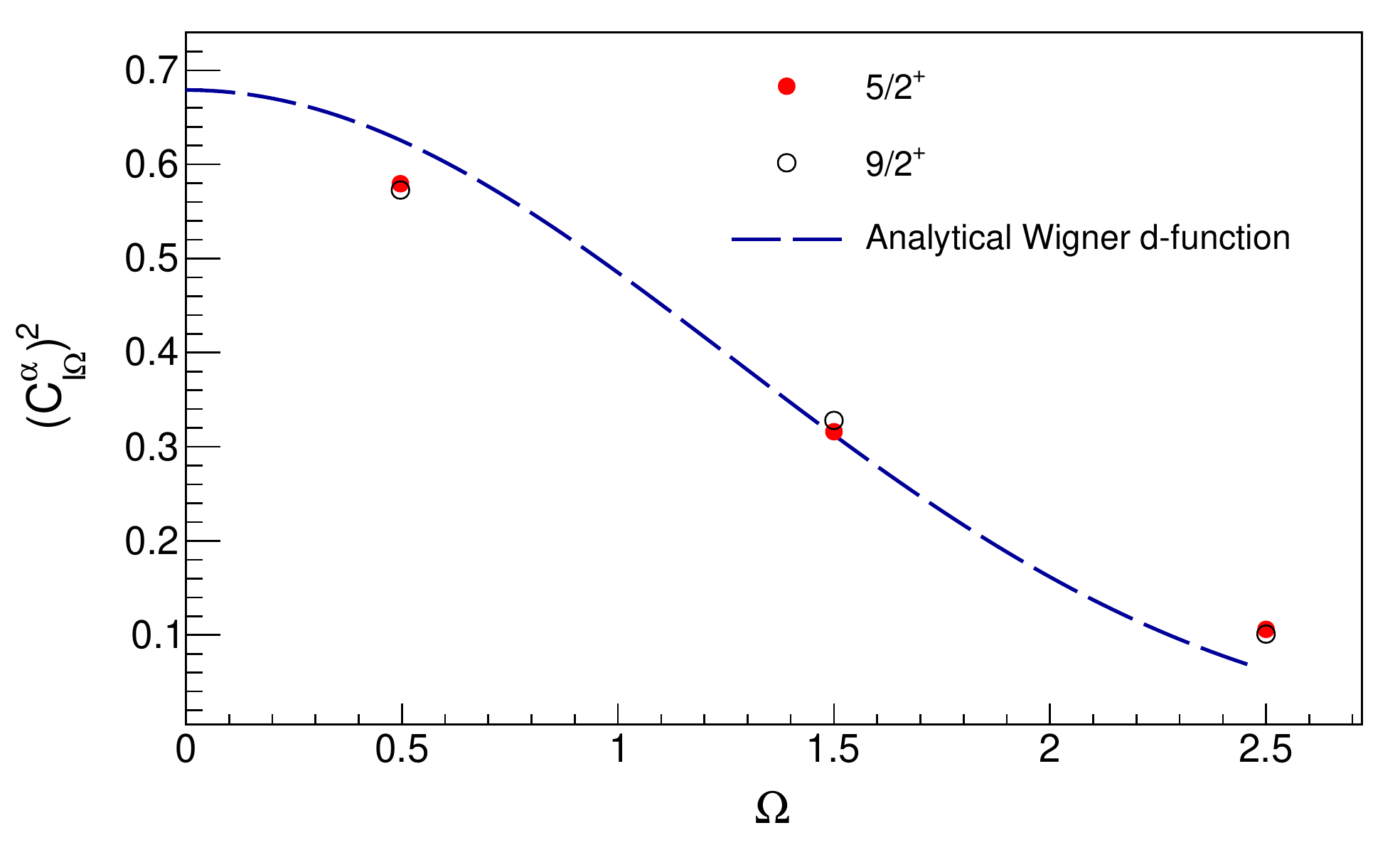}
\caption{ Wavefunctions of the $5/2^+$ (solid circles) and $9/2^+$ (open circles) states compared to the decoupled limit given by the $d$-function (dashed-line). }
\end{figure}

\noindent
If the conditions for a rotational aligned coupling scheme persist also in $^{30}$F, the odd-neutron will occupy the $\frac{1}{2}[330]$ level of the $f_{7/2}$  Nilsson multiplet.  In complete analogy with the odd-A case, 
we expect a doubly-decoupled band structure~\cite{Kreiner} with spins $I=(j_\pi+j_\nu), ~(j_\pi+j_\nu)+2, ~....   $, also following the core spacings.  Thus,  we predict the ground state of $^{30}$F
to be $6^-$. Of course, it is also possible that the occupation of the deformation driving $\frac{1}{2}[330]$ neutron level may polarize the core to a larger $\epsilon_2$ and a strong coupling scheme 
be realized, with the  $\Omega_\pi+\Omega_\nu $ configuration being favored and the $|\Omega_\pi-\Omega_\nu |$ nearby. In this limit the ground state will be $2^-$. \added{With the substantial difference in spin predicted for the ground state in these two coupling schemes, a measurement of the (unbound) ground-state resonance in $^{30}$F will be interesting to illuminate our understanding.}
\\

In summary,  the recent experimental results of Ref.~\cite{Pieter} and Shell Model calculations suggest the extension of the $N=20$ Island of Inversion to the Fluorine isotopes.
We have shown that the low-lying excitation spectrum of $^{29}$F can be understood in terms of a collective picture, with a level structure corresponding to the
rotation-aligned coupling limit of the PRM.   The Coriolis coupling effects on the proton $d_{5/2}$ Nilsson multiplet give rise to a (favored) decoupled band.  Thus, 
the $5/2^+$ bandhead naturally emerges as the ground state, and the $1/2^+$ as the first excited state, with its excitation energy depending directly on the $E_{2^+}(core)$.   
We have found a consistent solution at a deformation of 
 $\epsilon_2 \approx 0.17^{+0.15}_{-0.2}$ that suggests an excitation energy of the $2^+$ in $^{28}$O at $E_{2^+} \approx 2.\replaced{4}{3}\replaced{\pm0.2}{^{+0.2}_{-0.1}}$~MeV  in line with the conclusions reached
in Ref.~\cite{Pieter} based on the SDPF-M effective interaction.  PRM predictions for some spectroscopic observables were also presented.  Similar conditions in $^{30}$F would give rise to a
$\pi d_{5/2} \otimes \nu f_{7/2}$ double-decoupled structure with a predicted   $6^-$ ground state.   

%\begin{acknowledgments}
\section*{Acknowledgments}
 We would like to thank Frank Stephens for enlightening discussions and comments on the manuscript.
 This material is based upon work supported by the U.S. Department of Energy, Office of Science, Office of Nuclear Physics under Contract No. DE-AC02-05CH11231 (LBNL).

 %\end{acknowledgments}

\section*{References}

%\bibliography{mybibfile}

\end{document}